\documentclass[final,5p,times,twocolumn]{elsarticle}

\usepackage{graphicx}

\usepackage{amssymb}

\journal{Physica E}

\begin{document}

\begin{frontmatter}



\title{Measurement of diffusion thermopower in the quantum Hall systems}


\author{K. Fujita, A. Endo, S. Katsumoto and Y. Iye}

\address{Institute for Solid State Physics, University of Tokyo, Kashiwa, Chiba, 277-8581, Japan}

\begin{abstract}
We have measured diffusion thermopower in a two-dimensional electron gas at low temperature ($T$=40 mK) in the field range 0 $<B<$ 3.4 T, by employing the current heating technique. A Hall bar device is designed for this purpose, which contains two crossing Hall bars, one for the measurement and the other used as a heater, and is equipped with a metallic front gate to control the resistivity of the areas to be heated. In the low magnetic field regime ($B\leq$ 1 T), we obtain the transverse thermopower $S_{yx}$ that quantitatively agrees with the $S_{yx}$ calculated from resistivities using the generalized Mott formula. In the quantum Hall regime ($B\geq$ 1T), we find that $S_{yx}$ signal appears only when both the measured and the heater area are in the resistive (inter-quantum Hall transition) region. Anomalous gate-voltage dependence is observed above $\sim$1.8 T, where spin-splitting in the measured area becomes apparent. 
\end{abstract}

\begin{keyword}


Diffusion thermopower \sep Quantum Hall effect \sep Two-dimensional electron gas
\end{keyword}

\end{frontmatter}


\section{Introduction}
\label{Introduction}
The thermopower of a two-dimensional electron gas (2DEG) \cite{Fletcher86,Gallagher92,Fletcher99,Goswami09,Chickering09} has been attracting interest not only as a route to access its thermodynamic properties but also as a sensitive tool to probe various quantum phenomena that take place in a quantizing magnetic field (see, e.g., \cite{Yang,Granger}). The thermopower in a 2DEG contains contributions from two separate mechanisms: diffusion and phonon drag. It is well known that the latter is by far the dominant contribution in standard experiments using an external heater to introduce temperature gradient \cite{Fletcher99}. This is because the heater raises both the lattice and the electron temperatures alike; the heat current is thus predominantly carried by phonons, which generates the phonon-drag thermovoltage through the electron-phonon interaction. However, it is the diffusion thermopower that is expected to be more sensitive to the phenomena taking place in a 2DEG\@. Furthermore, the experimental results for diffusion thermopower will be much easier to interpret, since external complications, the phonons, are not involved. Therefore, it is desirable to have a method sensitive only to the diffusion contribution. This can be achieved by employing current heating technique, which induces gradient only in the electron temperature $T_{\rm e}$, leaving the lattice temperature intact. The technique was applied to a micro-scale (4$\times$8 $\mu m^2$) Hall bar by Maximov \textit{et al.} \cite{Maximov} to obtain diffusion contribution to the  longitudinal ($S_{xx}$) and transverse ($S_{yx}$) thermopower in the low magnetic field regime $B \leq 1.2$ T at a temperature $T$=1.6 K\@. Their use of the micro-Hall bar, however, resulted in rather large slowly-varying background attributable to the quasiballistic motion of electrons. In the present paper, we describe our attempt to acquire diffusion thermopower at dilution-refrigerator temperatures $\sim$ 40 mK, using a Hall bar designed to be well suited for the measurement of the thermopower, and having dimensions larger than the mean-free path of the electrons to avoid the intervention by the ballistic electrons. We make an attempt to extend the measurement to the quantum Hall regime, $B \geq 1$ T, employing a Hall bar device equipped with a front gate on the section used as a heater to circumvent the problem (to be discussed below) encountered in the quantum Hall regime.

\section{Sample and measurement method}
\label{Sample and measurement method}
\begin{figure}[!b]
\begin{center}
\includegraphics[width=\linewidth, clip]{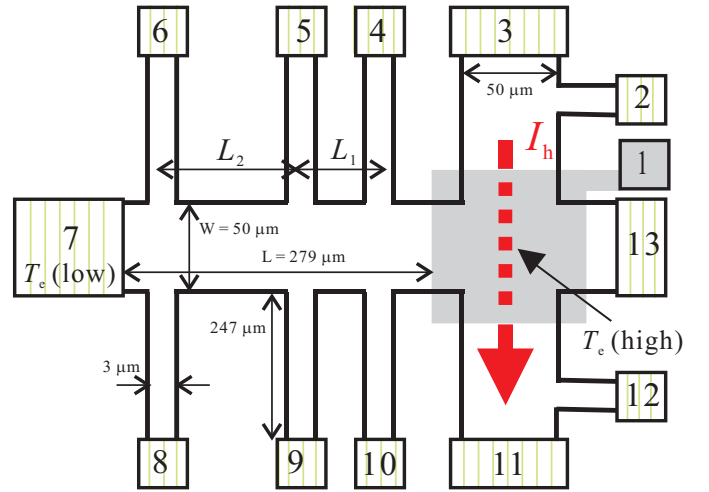}
\end{center}
\caption{
Schematic diagram of the sample. Hatched rectangles are the ohmic contacts. Main (horizontal, from 7 to 13) Hall bar (50$\times$279 $\mu$m$^2$) contains three pairs of voltage probes (4--6, 8--10). Secondary (vertical, from 3 to 11) Hall bar (170$\times$50 $\mu$m$^2$) is employed as the heater; the voltage probes (2,11) are used for the SdH measurement. A (gray) shaded rectangle is a metallic front gate to control the resistance of the area to be heated.}
\label{Fig1}
\end{figure}

A conventional GaAs/AlGaAs 2DEG wafer with the carrier density and mobility $n_{\rm e}$ $=$ $4 \times 10^{15}$ m$^{-2}$ and $\mu$ $=$ 70 m$^{2}$/(Vs), respectively, is patterned into the device geometry illustrated in Fig.\ \ref{Fig1}. The device is composed of two crossing Hall bars. The main Hall bar (between ohmic contacts 7 and 13) has a width $W$=50 $\mu$m and length $L$= 279 $\mu$m and contains three sets of the voltage probes (with contacts 4--6 and 8--10) to measure the longitudinal ($V_{xx}$) and transverse ($V_{yx}$) voltages at three different locations (or with different inter-probe distances for $V_{xx}$). Both $W$ and $L$, and distances between voltage probes ($L_1$=23 $\mu$m or $L_2$=153 $\mu$m), are designed to be much larger than the mean-free path $L_\mathrm{mfp}$=7.3 $\mu$m of the electrons.  The secondary Hall bar (between contacts 3 and 11), 170 $\mu$m-long and 50$\mu$m-wide, is used as a heater by driving an ac heating current $I_\mathrm{h}$=50--200 nA, with frequency $f$=13 Hz. The current $I_\mathrm{h}$ used is much larger than that in the ordinary resistivity measurement ($I$=0.5--10 nA) and raises the electron temperature $T_\mathrm{e}$ through Joule heating, but is kept small enough to prevent the heating of the lattice. We can probe the electron temperature in the heater section by the voltage probes (contacts 2 and 12), exploiting the amplitude of the Shubnikov-de Haas (SdH) oscillation. Thus the difference in the electron temperature $\Delta T_\mathrm{e}$=$T_{\rm e}$(high)$-T_{\rm e}$(low) is introduced to the main Hall bar, between $T_\mathrm{e}$(high) at the crossing region shared with the secondary Hall bar (dubbed as ``heater area'' henceforth) and $T_\mathrm{e}$(low) at the ohmic contact (contact 7) composed of diffused NiAuGe alloy, the latter assumed to be in equilibrium with the lattice temperature, or the temperature of the mixing chamber of the dilution refrigerator. Strictly speaking, ohmic contacts for the voltage probes (contact 4--6, 8--10) are also at $T_\mathrm{e}$(low), and therefore the temperature gradient is also directed from the heater area toward these contacts, resulting in rather complicated temperature distribution. To minimize the disturbance by the voltage-probe contact pads, arms are designed to be thin (3 $\mu$m) and long (247 $\mu$m), and the pads are made much smaller than that of contact 7 to diminish their efficiency as the heat sink.

Since $\Delta T_\mathrm{e} \propto I_\mathrm{h}^2$, thermopower $S_{xx}$ and $S_{yx}$ are obtained by detecting the component of $V_{xx}$ and $V_{yx}$ having the frequency $2f$=26 Hz by using a lock-in amplifier; then we have $S_{xx}=(V_{xx}/\Delta T_\mathrm{e})(L/L_{1(2)})$ and $S_{yx}=(V_{yx}/\Delta T_\mathrm{e})(L/W)$. The component of voltages with the frequency $f$, on the other hand, yields non-local resistance. We note in passing that since we are using a Hall bar, we can, of course, also measure the longitudinal and the Hall resistances (which can readily be translated to the resistivities $\rho_{xx}$ and $\rho_{yx}$) for the same area of the sample as the thermovoltages are acquired, simply by passing (small) current between contacts 7 and 13. 

With the method described so far, we succeeded in measuring the transverse (Nernst) component $S_{yx}$ of the diffusion thermopower for low magnetic fields $B\leq1$ T, as will be shown in the next section. The measurement of the longitudinal (Seebeck) component $S_{xx}$ is still suffering from the effect of electron deflection due to the magnetic field, which causes the mixing-in of the transverse component asymmetrically between $B>0$ and $B<0$. We therefore focus on the $S_{yx}$ component, acquired by using contacts 4 and 10 as voltage probes, in the present paper. A problem arises in applying the current heating technique in a magnetic field: since the longitudinal resistance varies with magnetic field also in the heater area, the temperature difference $\Delta T_\mathrm{e}$ generated by the Joule heating also varies with magnetic field. This does not cause serious trouble in low magnetic fields where the resistance variation is not so large. In the quantum Hall regime, however, resistance variation is phenomenal, ranging from $\sim$0 at the quantum Hall states to $\sim$k$\Omega$ in between. To avoid the difficulties, we placed a metallic front gate on the heater area, as shown in Fig.\ \ref{Fig1}. This enabled us to control the carrier density, hence the resistance, of the heater area independent of whether or not the measured area is in the quantum Hall states.

\section{Results}
\label{Results}
\begin{figure}[tb]
\includegraphics[width=\linewidth, clip]{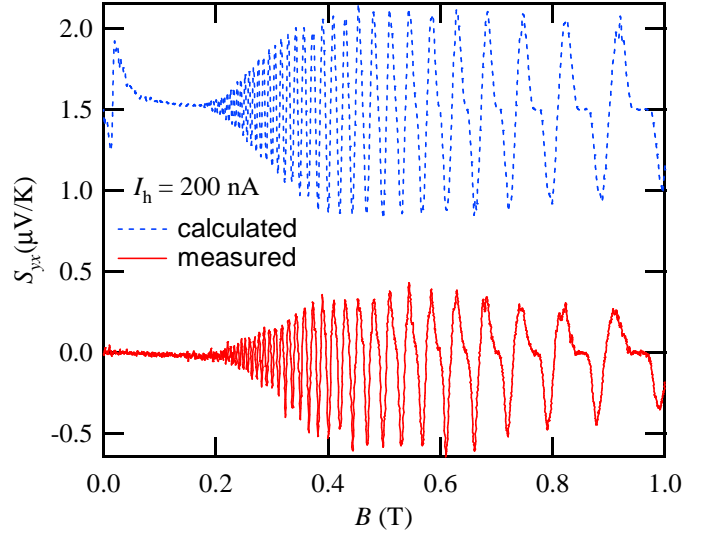}
\caption{
Transverse thermopower $S_{yx}$ measured directly (solid line) and that calculated from the measured $\rho_{xx}$ and $\rho_{yx}$ using Eq.\ (\ref{eq:nernstB}) (dotted line), at $T$ $=$ 40 mK. The latter is offset by 1.5 $\mu$V/K for clarity.}
\label{Fig2}
\end{figure}

Figure.\ \ref{Fig2} shows the transverse thermopower $S_{yx}$ at low magnetic fields ($B \leq 1$ T) measured by the method described in the previous section, with a heating current $I_\mathrm{h}$=200 nA\@. (A sample without the metallic front gate is used for this measurement.) Note that $S_{yx}$ oscillates around zero, without any noticeable background. The diffusion thermopower is related to the longitudinal and transverse conductivities $\sigma_{xx}$ and $\sigma_{yx}$ by the generalized Mott formulas \cite{Jonson84}, 
\begin{equation}
S_{xx}=-L_0eT\frac{d}{d\varepsilon_\mathrm{F} }\ln \sqrt{\sigma_{xx}^2+\sigma_{yx}^2}, \label{eq:seebeckE}
\end{equation}
\begin{equation}
S_{yx}=-L_0eT\frac{d}{d\varepsilon_\mathrm{F} }\arctan \frac{\sigma_{yx}}{\sigma_{xx}}, \label{eq:nernstE}
\end{equation}
where $L_0 = \pi^2{k_\mathrm{B}}^2/3e^2$ is the Lorenz number and $\varepsilon_\mathrm{F}$ the Fermi energy.  
If we assume that properties of the system are mainly determined by the location of the Fermi energy with respect to the Landau levels, we can identify the energy derivative with the derivative with respect to the magnetic field as
\begin{equation}
\frac{d}{d\varepsilon_\mathrm{F}}=-\frac{B}{\varepsilon_F}  \frac{d}{dB}\label{eq:EtoB}.
\end{equation}
Using this relation, Eqs.\ (\ref{eq:seebeckE}) and (\ref{eq:nernstE}) are rewritten as
\begin{equation}
S_{xx}=\frac{L_0eTB}{E_F} \frac{d}{dB }\ln \sqrt{\rho_{xx}^2+\rho_{yx}^2} \label{eq:seebeckB}
\end{equation}
\begin{equation}
S_{yx}=\frac{L_0eTB}{E_F} \frac{d}{dB }\arctan \frac{\rho_{yx}}{\rho_{xx}}, \label{eq:nernstB}
\end{equation}
where we replaced the conductivities by the resistivities by inverting the tensor. We can thus calculate $S_{xx}$ and $S_{yx}$ using  $\rho_{xx}$ and $\rho_{yx}$ measured in the same Hall bar. In Fig.\ \ref{Fig2}, we also plot $S_{yx}$ calculated by Eq.\ (\ref{eq:nernstB}). The good quantitative agreement between the two traces confirms that the measured $S_{yx}$ is actually derived from the diffusion contribution. Discrepancy noted at very low magnetic fields ($< \sim$0.05 T) can be traced back to the negative magnetoresistance in $\rho_{xx}$, presumably originating from the electron interactions \cite{Li03}. The absence of the similar effect in the thermopower is intriguing, but the reason is currently unknown.  
\begin{figure}[tb]
\includegraphics[width=\linewidth, clip]{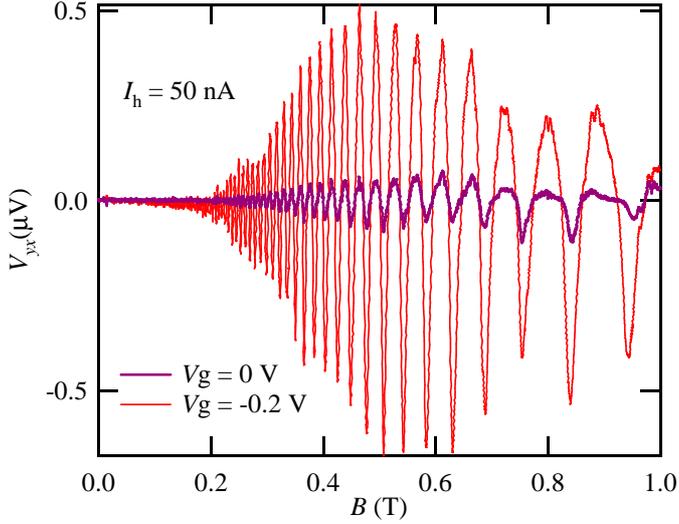}
\caption{
Transverse thermovoltages $V_{yx}$ measured with two different gate voltages $V_\mathrm{g}$, with the same heating current $I_\mathrm{h}$= 50 nA. $T$ $=$ 40 mK\@.}
\label{Fig3}
\end{figure}

Next, we check the function of the front gate. In Fig.\ \ref{Fig3}, we plot traces of the transverse thermovoltage $V_{yx}$ measured with two different values of the gate voltages $V_\mathrm{g}$, using the same heating current $I_\mathrm{h}$=50 nA\@. With more negative $V_\mathrm{g}$, resistivity in the heater area increases accompanying the decrease in the electron density. The heater area therefore achieves higher electron temperature with the same $I_\mathrm{h}$, resulting in larger amplitude in $V_{yx}$. The two traces virtually overlap each other, when translated into $S_{yx}$ by using measured values of $\Delta T_\mathrm{e}$. 

\begin{figure}[!b]
\includegraphics[width=\linewidth, clip ]{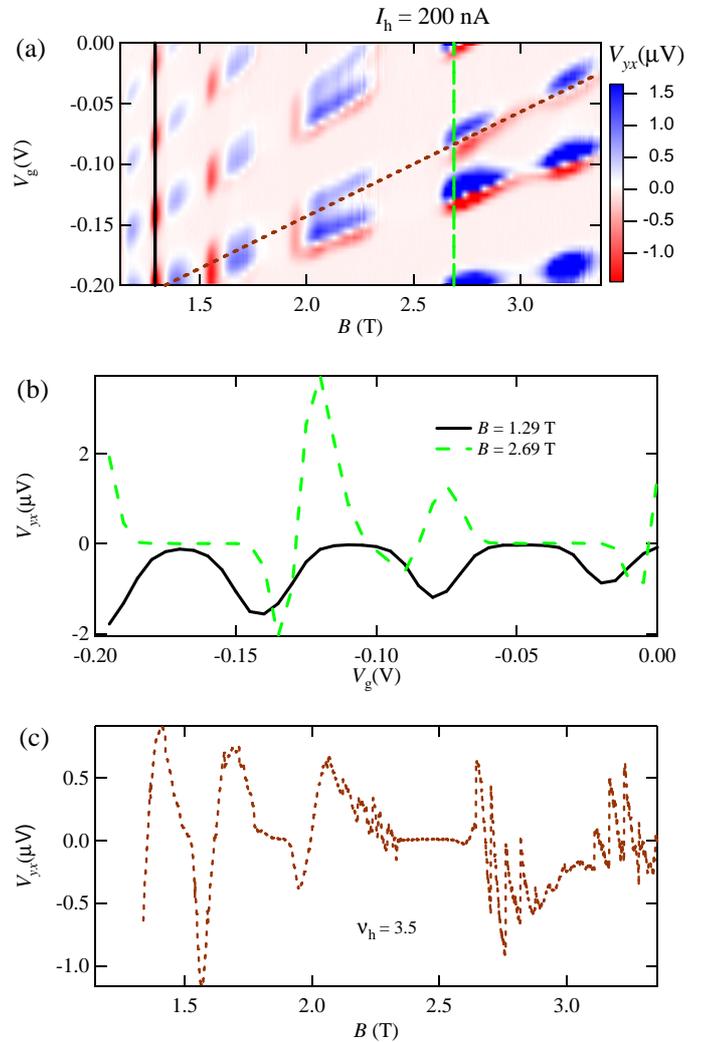}
\caption{
(a) Color-scale intensity plot of the transverse thermovoltage $V_{yx}$ as a function of $B$ and $V_\mathrm{g}$. $I_\mathrm{h}$= 200 nA\@. (b) Cross sections of (a) at fixed magnetic fields, plotted as a function of $V_g$. Solid line: $B$=1.29 T\@.  Dashed line: $B$ $=$ 2.69 T\@. (c) Cross section of (a) with a fixed filling factor $\nu_\mathrm{h}$= 3.5 of the heater area, plotted as a function of $B$. Cross sections in the $B$-$V_\mathrm{g}$ plane are indicated in (a) by the same line-type as in (b) and (c).}
\label{Fig4}
\end{figure}

We now move on to the quantum Hall regime. Fig.\ \ref{Fig4} (a) shows color-scale intensity plot of $V_{yx}$ in the $B$-$V_\mathrm{g}$ plane. The plot was obtained by the repetitive $B$-sweeps with fixed $V_\mathrm{g}$, varying the $V_\mathrm{g}$ step by step with the increment of 0.005 V\@. Note that $V_\mathrm{g}$ alters the filling factor $\nu_\mathrm{h}$ of the heater area for a fixed $B$; the region of the sample to be measured is not affected by $V_\mathrm{g}$. Finite  signal of $V_{yx}$ appears only when both the measured area and the heater area are in the resistive (the inter-quantum-Hall transition) regime; temperature gradient is not generated unless the heater area possesses a finite resistivity, and $S_{yx}$ equals zero for $\rho_{xx}$=0 and a finite value of $\rho_{yx}$ in the measured area (see Eq.\ (\ref{eq:nernstB})). 
It can be seen that the plot considerably changes its appearance below and above $B\sim1.8$ T\@. Below $\sim$1.8 T, the $V_{yx}$ signal is basically determined by $B$ and does not depend much on $V_\mathrm{g}$ as long as the heater area is in the resistive regime. Above $\sim$1.8 T, by contrast, $V_{yx}$ exhibits rather complicated pattern dependent both on $B$ and $V_\mathrm{g}$. The difference is more apparent in the cross sections, shown in Fig.\ \ref{Fig4}(b), at fixed values of $B$. For $B< \sim$1.8 T, the cross section shows trains of dips (or peaks, depending on the value of $B$), with the depth slightly varying with $V_\mathrm{g}$ attributable to the slight difference in the resistivity of the heater area. For $B> \sim$1.8 T, on the other hand, $V_{yx}$ alternates sign with $V_\mathrm{g}$, with the sign reversal taking place roughly at the half-filling of the Landau levels in the heater area, namely at $\nu_\mathrm{h} \sim$ half integer. The latter is a rather anomalous behavior that defies simple interpretation, since, as noted above, $V_\mathrm{g}$ affects only the heater area, leaving the measured area intact. We currently have no clear explanation for this observation. A possible origin is the non-uniform heating of the heater area in the quantum Hall regime. It has been shown, both experimentally \cite{Komori} and theoretically \cite{Ise}, that a 2DEG subjected a large current develops a distribution in the electron temperature in the quantum Hall regime owing to the Ettingshausen effect. The distribution varies rapidly with the filling factor; the hot and cold region alternates roughly at integer and half integer filling factors \cite{Ise}. Apparently, detailed knowledge of the temperature distribution both in the heater area and the measured area is necessary for the interpretation of the measured thermovoltage.
We would also like to point out that the border field $B \sim$1.8 T coincides with the onset of the spin splitting in the measured area, suggesting the possible involvement of the spins in the anomalous behavior. 

To avoid the complications arising from the behavior of the heater area, we look at the data at fixed $\nu_\mathrm{h}$. A cross section of Fig.\ \ref{Fig4} (a) at $\nu_\mathrm{h}$= 3.5 is shown in Fig.\ \ref{Fig4} (c). The trace is more or less the extension of the low-field traces, Figs. \ref{Fig2} and \ref{Fig3}, anticipated for the diffusion contribution to the $S_{yx}$, albeit with rather large noise in the high-field regime.

\section{Conclusion}
\label{Conclusion}
We have measured diffusion contribution to the transverse thermopower $S_{yx}$, by using a specially designed Hall bar device having dimensions much larger than the mean-free path of the electrons. We obtain good quantitative agreement with $S_{yx}$ calculated from the resistivities using the generalized Mott formula Eq.\ (\ref{eq:nernstB}) in the low magnetic field regime ($B\leq$ 1T).
In the quantum Hall regime ($B\geq$ 1T), we find, by mapping out the transverse thermovoltage $V_{yx}$ in the $B$-$V_\mathrm{g}$ plane, that $V_{yx}$ signal appears only when both the measured and the heater area are in the resistive regime and that $V_{yx}$ exhibits rather anomalous $V_\mathrm{g}$ dependence above $\sim$1.8 T\@. The results suggest that the diffusion thermopower can be obtained also in the quantum Hall regime by picking out the $V_{yx}$ at the constant filling factor $\nu_\mathrm{h}$ at the heater area.

\section*{Acknowledgment}
A. E. acknowledges the financial support by Grant-in-Aid for Scientific Research (B) (20340101) from the Ministry of Education, Culture, Sports, Science and Technology (MEXT) and by the National Institutes of Natural Sciences undertaking Forming Bases for Interdisciplinary and International Research through Cooperation Across Fields of Study and Collaborative Research Program (No.~NIFS08KEIN0091).



\end{document}